\documentclass{emulateapj}
\usepackage{apjfonts}
\usepackage{verbatim}
\usepackage{epsfig}

\shorttitle{Detonations in Sub-Chandrasekhar Mass C+O White Dwarfs}
\shortauthors{Sim et al.}

\begin{document}

\title{Detonations in Sub-Chandrasekhar Mass C+O White Dwarfs}

\author{S. A. Sim, F. K. R\"{o}pke, W. Hillebrandt,
  M. Kromer, R. Pakmor, M. Fink, A. J. Ruiter, I. R. Seitenzahl}
\affil{Max-Planck-Institut f{\"u}r Astrophysik,
  Karl-Schwarzschild-Str.~1, D-85748 Garching, Germany}

\begin{abstract}
Explosions of 
sub-Chandrasekhar-mass white dwarfs are one alternative to the
standard Chandrasekhar-mass model of Type Ia supernovae. They are
interesting since binary systems with sub-Chandrasekhar-mass primary
white dwarfs should be common and this scenario would suggest a simple
physical parameter which determines the explosion brightness, namely
the mass of the exploding white dwarf. Here we perform one-dimensional
hydrodynamical simulations, associated post-processing
nucleosynthesis and multi-wavelength radiation transport calculations
for pure detonations of carbon-oxygen white dwarfs.
The
light curves and spectra we obtain from these simulations are
in good agreement with observed properties of Type Ia
supernovae. 
In particular, for white dwarf masses from 0.97--1.15~M$_{\odot}$ we obtain $^{56}$Ni masses between 0.3 and
0.8~M$_{\odot}$, sufficient to capture almost the complete range of
Type Ia supernova brightnesses. 
{Our optical light curve rise times, peak colours and decline
timescales
display trends which are generally consistent with observed
characteristics although the range of $B$-band decline timescales
displayed by our current set of models is somewhat too narrow.} 
In agreement with observations, the maximum light spectra of the
models show clear features associated with intermediate mass elements
and reproduce the sense of the observed correlation between 
explosion luminosity and 
the
ratio of the Si~{\sc{ii}} lines at $\lambda6355$ and $\lambda5972$.
We therefore suggest that sub-Chandrasekhar mass explosions
are a viable model for Type Ia supernovae for any binary evolution
scenario {leading to explosions} in which the optical display is dominated by
the material produced in {a detonation} of the primary white dwarf.
\end{abstract}


\section{Introduction}

In recent years, considerable work has been devoted to the study
of the 
Chandrasekhar-mass ($M_\mathrm{Ch}$) model of Type~Ia supernovae
(SNe~Ia). As shown by \cite{arnett1971a}, prompt detonations of
$M_\mathrm{Ch}$ carbon/oxygen (C+O) white dwarfs (WDs) in hydrostatic
equilibrium mainly produce iron group elements (IGEs). Thus, they cannot
account for the significant amounts of 
intermediate-mass elements (IMEs; e.g.\ silicon and sulphur) 
responsible for the features which dominate the maximum light spectra. 
To obtain these, pre-expansion of the WD is necessary such that
burning partially takes place under low-density conditions where
IMEs can be synthesized.
One way of achieving this is provided by
models in which the flame ignites as a deflagration which releases
sufficient energy to expand the star before a
deflagration-to-detonation transition occurs 
{(\citealt{khokhlov1991a})}.
An
alternative to this pre-expansion is the detonation of a
sub-Chandrasekhar mass (sub-$M_\mathrm{Ch}$) WD starting from a hydrostatic
configuration. Here, a variety of density profiles can be realized,
determined by the WD mass. Close to $M_\mathrm{Ch}$, the detonation
produces primarily IGEs and few IMEs, while for less
massive WDs more IMEs and less IGEs
will be synthesized. 

Detonation of a sub-$M_\mathrm{Ch}$ WD cannot occur
spontaneously but must be triggered by external compression. 
The most widely discussed mechanism for sub-$M_\mathrm{Ch}$
explosions has been the \emph{double detonation} model. Here, a C+O~WD accretes from
a companion star and develops a 
helium-rich outer shell. 
This may occur for binaries with  
helium-rich donors or hydrogen-rich donors where the accreted
hydrogen is burned to helium.
If the helium-shell becomes sufficiently massive, it can become unstable and
detonate. Subsequent compression 
of the core by inward propagating shocks may produce a secondary carbon
detonation which explodes the WD (e.g. \citealt{woosley1986a,fink2007a}).
Detonations in helium-rich surface layers have also been discussed
for the case of rapid dynamical mass transfer in binary systems
containing a C+O~WD with a helium-rich WD companion
\citep{guillochon2009}. In that case instabilities in the
accretion seed dense knots which, by impacting on the underlying
WD surface, might trigger a detonation in the accreted helium
leading to a potential secondary core detonation. 
It has also been speculated
that sub-$M_\mathrm{Ch}$ explosions may arise during violent accretion
in mergers of C+O~WD binaries. Here, the C+O accretion may lead to an
edge-lit detonation or carbon flashes that trigger a
core detonation (see e.g.\ \citealt{shigeyama1992} but for a different
result see \citealt{loren2009b}).

Most previous work on testing sub-$M_\mathrm{Ch}$ models has
focused on cases in which the core detonation is triggered by
detonation in an overlying massive shell ($\sim0.2{\;}\mbox{M}_{\odot}$) of
helium
(e.g. \citealt{woosley1994b, livne1995a, hoeflich1996a,hoeflich1996b,nugent1997}). 
In those models burning in the helium shell synthesizes significant masses
of $^{56}$Ni in the outer ejecta, leading to spectra and
light curves in conflict with observations.
As noted in those studies, however, these conclusions are strongly dependent on the influence of the
shell material. In particular, they may not be
applicable if such a layer is absent (or much less massive) 
or if its post-burning composition lacks $^{56}$Ni.
Recently, \cite{bildsten2007a} suggested that
detonation of the helium-shell may be possible for a shell with mass
as low as $\sim0.055{\;}\mbox{M}_{\odot}$ 
around a $1.025{\;}\mbox{M}_{\odot}$ C+O core and that the burning produces only $0.012{\;}\mbox{M}_{\odot}$ of $^{56}$Ni along with some lighter
IGEs
(\citealt{guillochon2009} find that even
lower atomic-number burning products can dominate in their helium
detonations).
Even for the low shell masses of \cite{bildsten2007a}, \cite{fink2010}
find that a secondary core detonation is possible.
To date, sub-$M_\mathrm{Ch}$ explosions in the
absence of a nickel-rich outer layer have not been studied in 
detail. \cite{shigeyama1992} investigated the explosion
dynamics of sub-$M_\mathrm{Ch}$ detonations and concluded that their
 characteristic properties were consistent with SNe~Ia but
they did not perform realistic radiative transfer simulations.

A full treatment of any class of sub-$M_\mathrm{Ch}$ explosion model
requires realistic hydrodynamical and nucleosynthesis simulations of the
accretion phase,
triggering mechanism and subsequent explosion.
Here, however, we present a simple numerical experiment that is
relevant to any class of sub-$M_\mathrm{Ch}$ explosion model. We consider pure
detonations of
sub-$M_\mathrm{Ch}$ C+O~WDs with different masses, neglecting the   
question of how this detonation is initiated. 
This allows us to investigate the idealized case of 
sub-$M_\mathrm{Ch}$ detonation scenarios in which the observable display is
dominated by material produced in the core explosion.
Our goal is to determine the extent to which the least ambiguous
component of the system, namely the detonation of a 
sub-$M_\mathrm{Ch}$ C+O~WD, could lead to explosions which are 
consistent with
observations of SNe~Ia.

\section{Hydrodynamic Simulations}

To explore the properties of detonations of
sub-$M_\mathrm{Ch}$ WDs, we set up five {hydrostatic} models with WD
masses $(M_\mathrm{WD})$
ranging from 0.81 to $1.15{\;}\mbox{M}_{\odot}$ (see Table~\ref{tab:RT_out}) 
with centrally-ignited detonations.  
{We performed the simulations in an axisymmetric setup of
the full star but as the initial conditions were spherically symmetric
and the symmetry is preserved during evolution, our models are
one-dimensional.
The
simulations were performed with our SNe~Ia explosion code (the initial
sizes of the computational grid cells are given in
Table~\ref{tab:RT_out}; see
\citealt{fink2010} for technical details).} 

The detonations were represented with the level-set technique
\citep{reinecke1999a}. This requires as inputs the detonation velocity 
and the energy release in the burning. 
The detonation velocities take into account pathological
detonation speeds at high fuel densities \citep{gamezo1999a} whereas
low-density detonations are
assumed to be of Chapman-Jouguet type {(see \citealt{fink2010} for
details)}.
The energy released by nuclear burning
($E_\mathrm{nuc}$) and the 
asymptotic kinetic energy of the ejecta ($E_\mathrm{k}$) are given for each
model in Table~\ref{tab:RT_out}.

\begin{table*}
\begin{center}
\caption{Nucleosynthesis products and optical light curve properties$^a$ for detonations of
  white dwarfs.} 
\begin{tabular}{lllllll} \hline
Parameter & \multicolumn{6}{c}{Models} \\ \hline
$M_\mathrm{WD}^b$ (M$_{\odot}$) & 1.15 & 1.06 & 1.06 & 0.97 & 0.88 & 0.81\\
$\rho_{c}^c$ (g~cm$^{-3}$)& $7.9\times 10^7$ & $4.15\times 10^7$ & $4.15\times 10^7$ & $2.4\times 10^7$ & $1.45\times 10^7$ & $1.0\times 10^7$\\
WD comp. (C/O/Ne)$^d$& 50/50/0 & 50/50/0 & 42.5/50/7.5 &50/50/0  & 50/50/0 & 50/50/0 \\
$E_\mathrm{nuc}^e$ (foe)& 1.64 & 1.41 & 1.41 & 1.19 & 0.97 & 0.82\\
$E_\mathrm{k}^e$ (foe)& 1.39 & 1.22 & 1.22 & 1.04 & 0.86 & 0.73\\
Cell size$^{f}$ (cm) & $1.0 \times 10^6$ & $1.1 \times
10^6$ & $1.1 \times 10^6$ & $1.3 \times 10^6$ & $1.5 \times 10^6$ & $1.7 \times 10^6$ \\
$M_{^{56}\mbox{\scriptsize Ni}}^g$ (M$_{\odot}$)& 0.81 & 0.56 & 0.43
& 0.30 & 0.07 & 0.01\\
$M_\mathrm{IGE}^g$ (M$_{\odot}$)& $2.1\times 10^{-2}$&$1.7\times 10^{-2}$&$1.8\times 10^{-1}$&$1.2\times 10^{-2}$&$4.7\times 10^{-3}$&$8.7\times 10^{-4}$\\
$M_\mathrm{IME}^g$ (M$_{\odot}$)&0.27 & 0.41 & 0.36 & 0.54 & 0.63 &
0.57\\
$M_\mathrm{O}^g$ (M$_{\odot}$)&0.04 & 0.08 &0.09 & 0.12 & 0.17 & 0.22\\
${\Delta}m_{15}$ (mag.)&1.34 & 1.56 & 1.42 & 1.73 & 1.77 & --\\
$t^\mathrm{B}_\mathrm{max}$ (mag.)&19.2 & 20.1 & 18.0 & 19.9 & 14.1& --\\
$B_\mathrm{max}^h$ (mag.)&-19.9 & -19.2 & -18.7 & -18.5 & -16.6& --\\
$V_\mathrm{max}^h$ (mag.)&-19.6 & -19.4 & -19.3 & -18.8 & -17.3& --\\
$R_\mathrm{max}^h$ (mag.)&-19.8 & -19.3 & -19.3 & -18.8 & -17.7& --\\
$I_\mathrm{max}^h$ (mag.)&-19.6 & -19.0 & -19.2 & -18.7 & -17.8& --\\
$(B - V)_\mathrm{max}^h$ (mag.)&0.15 & 0.13 & 0.48 & 0.24 & 0.63 &
--\\
$v_{\mathrm{Si II}}^i$ (km~s$^{-1}$)\rule[-0.15cm]{0.0cm}{0.0cm} & 12,500 & 11,500 &
11,500 & 9,000 & 6,000&--\\
 \hline
\end{tabular}
\end{center}
\vspace{-0.3cm}\hspace{2.4cm}\begin{minipage}{13cm}
{$^a$ Since it has very low $^{56}$Ni-mass, we did not perform
radiative transfer simulations for the 0.81~M$_{\odot}$ model.}\\
{$^b$ Mass of white dwarf.}\\
{$^c$ Central density of white dwarf.}\\
{$^d$ Initial composition of WD (percentage by mass of $^{12}$C/$^{16}$O/$^{22}$Ne)}\\
{$^e$ Energy released by nuclear burning ($E_\mathrm{nuc}$) and asymptotic
kinetic energy of the ejecta ($E_\mathrm{k}$).}\\
{{$^f$ Initial size of computational grid cells in the WD. Since the
simulations use an expanding grid, the physical resolution degrades with time.}}\\
{{$^g$ Mass yields for $^{56}$Ni, stable iron group elements (IGEs)},
intermediate-mass elements (IMEs) and oxygen (O).}\\
{$^h$ Peak magnitudes are given at the true peaks in each band. Colours
are quoted at time ($t^\mathrm{B}_\mathrm{max}$) of $B$-band maximum.}\\
{$^i$ Blueshift velocity of Si~{{\sc}ii}~$\lambda$6355 at
$t^\mathrm{B}_\mathrm{max}$.}\\
\end{minipage}
\label{tab:RT_out}
\end{table*}

\section{Nucleosynthesis calculations}

The nucleosynthesis was computed using our 
standard tracer-particle technique \citep{travaglio2004a}.
The masses obtained for $^{56}$Ni, stable IGEs, IMEs and O are given in Table~\ref{tab:RT_out}.

We first computed nucleosynthesis for all five explosion simulations
assuming an initial WD composition of uniformly mixed
$^{12}$C and $^{16}$O with equal mass fractions {(``pure-C+O'', hereafter)}. 
Although commonly adopted in Type~Ia explosion
modelling, this composition is not strictly correct since the mass-fraction of
$^{16}$O is expected to be larger than $^{12}$C in the inner regions
(e.g. \citealt{salaris1997}).
For $M_\mathrm{Ch}$ delayed-detonation models, 
\cite{dominguez2001a} showed that the C/O-ratio affects the
$^{56}$Ni-mass by {$\sim14$\%} and the velocity
structure of the ejecta by up to a few $1000{\;}\mbox{km}{\;}\mbox{s}^{-1}$. Our
calculations are expected to have a similar sensitivity to the adopted
C/O-ratio.
{Moreover, C+O~WDs formed from progenitors with non-zero metallicity
will be polluted by some $^{22}$Ne.
Since the neutron excess of $^{22}$Ne significantly affects the
nucleosynthesis, for one of the hydrodynamical models
($M_\mathrm{WD}=1.06{\;}\mbox{M}_{\odot}$) we
repeated the nucleosynthesis post-processing step adopting a high initial
$^{22}$Ne mass fraction of 7.5\% (this would corresponds to a rather
high metallicity progenitor, $Z_{0}\sim$3$Z_{\odot}$).
This model (``C+O+Ne'', hereafter)
allows us to bracket some of the systematic uncertainties associated
with progenitor composition.}
The nucleosynthesis yields for this model are also given in Table~\ref{tab:RT_out}.

The range of $M_\mathrm{WD}$ we consider leads to $^{56}$Ni-masses 
from $\sim0.01{\;}\mbox{M}_{\odot}$ to $0.81{\;}\mbox{M}_{\odot}$, wide enough to
encompass the range implied for all but the brightest SNe~Ia
(see e.g. \citealt{stritzinger2006a}). 
The lowest mass models $M_\mathrm{WD}=0.81$ and $0.88{\;}\mbox{M}_{\odot}$
make very little $^{56}$Ni ($\sim$0.01 and $0.07{\;}\mbox{M}_{\odot}$,
respectively). 
Thus they would be faint and
lie outside the range of normal SNe~Ia. Throughout the following, we 
therefore neglect further discussion of the
$M_\mathrm{WD}=0.81{\;}\mbox{M}_{\odot}$ model but 
retain the $0.88{\;}\mbox{M}_{\odot}$ model as a
point of reference for the faintest observed SNe~Ia.

\begin{figure}
\hspace{-0.5cm}
\vspace{-0.3cm}
\epsfig{file=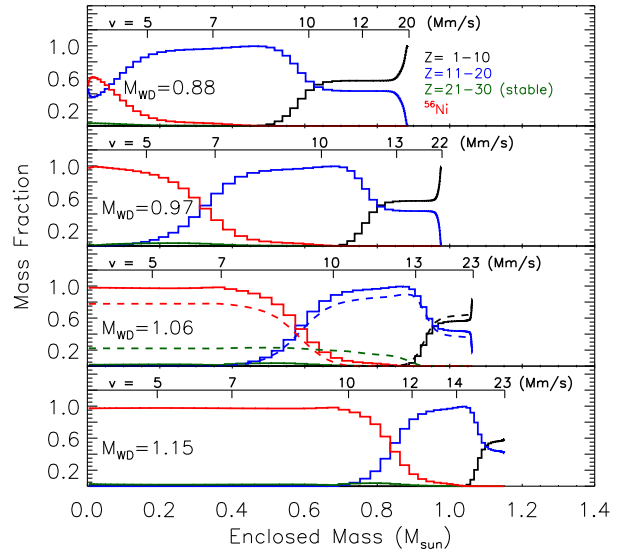,width=8cm,angle=90}
\caption{Composition versus mass-coordinate for
  the four models that produced significant masses of $^{56}$Ni 
  (least- to most-massive; top to bottom). The black
  histograms indicate low-mass elements, blue shows
  intermediate-mass elements ($Z=11-20$). Red represents
  $^{56}$Ni while stable iron group elements ($Z=21-30$)
   are shown in green. In the third panel, the solid lines show the
   results for the {pure-C+O model} while the
   dashed lines show the {C+O+Ne model}. In each
   panel the velocity scale (in Mm~s$^{-1}$) is also indicated.}
\label{fig:abund}
\end{figure}

Figure~\ref{fig:abund} shows the stratification of
the nucleosynthesis products grouped into low-mass elements, IMEs,
stable IGEs and $^{56}$Ni. The structure of the
models is very similar to that obtained by
\cite{shigeyama1992}: small masses of stable IGEs are produced and the mass-shell in which large
mass-fractions of IMEs are produced is fairly
extended. 
Significant IME mass-fractions are present up to almost the
highest velocities in all models, consistent with the lower
limits on the outer extent of Si-rich material discussed by
\cite{mazzali2007a}. Moreover, a clear trend exists whereby the inner
boundary of the IME-rich layers lies at higher velocities in the 
models where the $^{56}$Ni-mass is larger, the same trend as inferred
from observations \citep{mazzali2007a}. 

The most important consequence of $^{22}$Ne in our {C+O+Ne model} is
a substantial increase in the mass of stable IGEs (see
Table~\ref{tab:RT_out} and
third panel of Figure~\ref{fig:abund}; see also \citealt{hoeflich1998a}). These extend over a wide range
of mass-coordinate and come at the expense of less
$^{56}$Ni in the inner regions and fewer IMEs in the outer zones.

\section{Radiative Transfer simulations}

For each of the four detonation models that produce
$>0.05$~M$_{\odot}$ of $^{56}$Ni, we performed radiative transfer simulations using
our Monte Carlo code {\sc{artis}}
(\citealt{sim2007b,kromer2009a}). For
$M_\mathrm{WD}=1.06$~M$_{\odot}$, we ran simulations for
both our pure-C+O and C+O+Ne models.
For all calculations we used our largest atomic data set
($\sim8.2\times10^6$ lines)
and our NLTE
treatment of ionization (see \citealt{kromer2009a}).
Table~\ref{tab:RT_out} gives the light curve decline-rate
parameter (${\Delta}m_{15}$)\footnote{${\Delta}m_{15}$ is defined as the
  change in $B$-band magnitude between maximum light and 15~days
  thereafter.}, the time of $B$-band maximum light ($t^\mathrm{B}_\mathrm{max}$),
the optical peak magnitudes and the $B-V$ colour at
$t^\mathrm{B}_\mathrm{max}$.  

\section{Comparison with observations}

\begin{figure*}
\begin{center}
\epsfig{file=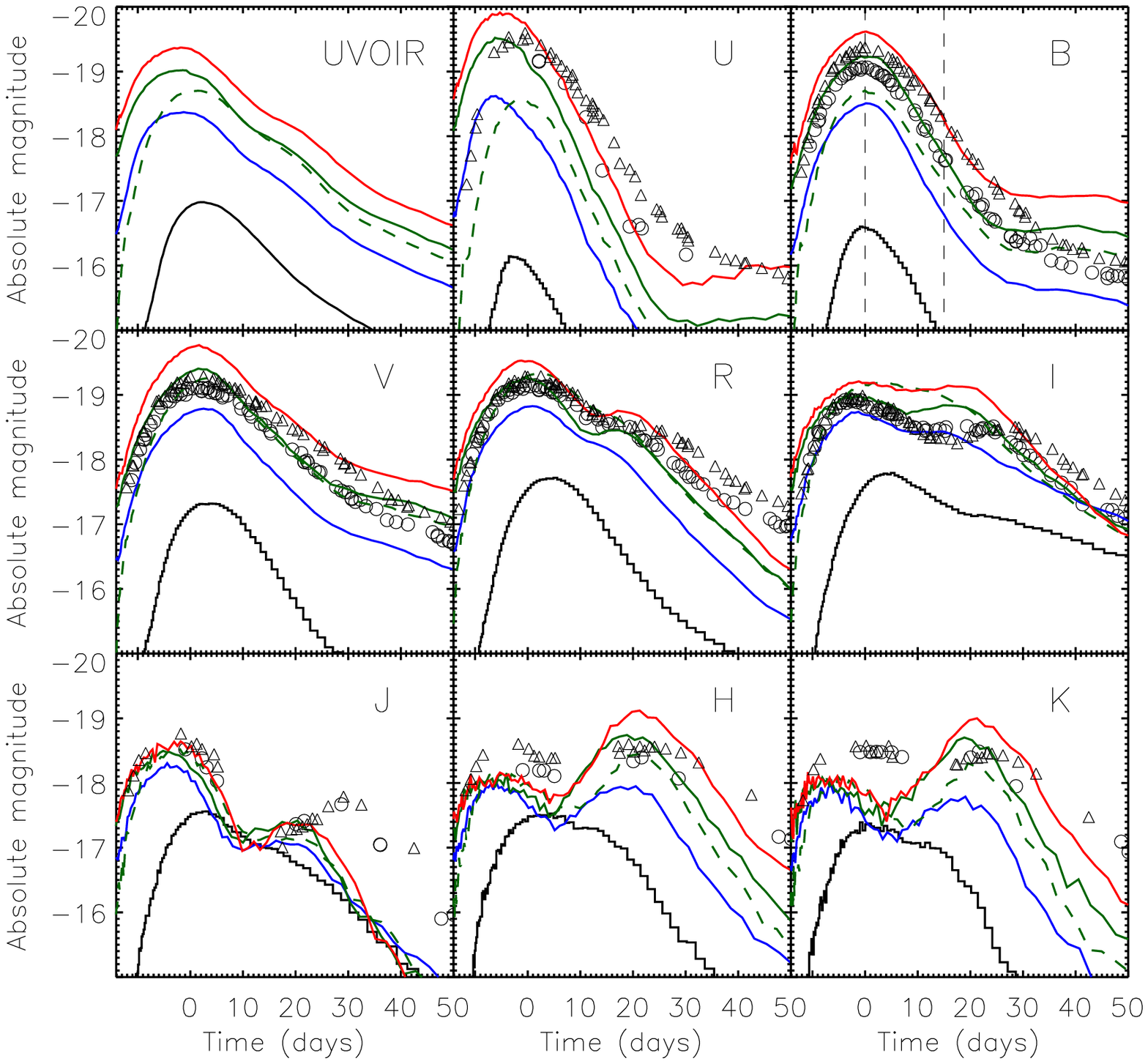,width=12cm,angle=90}
\end{center}
\caption{Computed \textit{UVOIR} bolometric, $U$, $B$, $V$, $R$, $I$, $J$, $H$
  and $K$-band light curves for $M_\mathrm{WD}=1.15$,
  1.06, 0.97, 0.88 M$_{\odot}$ (red, green, blue and black,
  respectively). 
For $M_\mathrm{WD}=1.06$~M$_{\odot}$, light curves for both our {pure-C+O}
and {C+O+Ne models} are
shown (solid and dashed green lines, respectively). 
Photometry for SN~2004eo 
(black circles; \citealt{pastorello2007b})
and SN~2005cf (black triangles; \citealt{pastorello2007a})
are also shown.
The observations have been corrected for reddening and 
  distance using parameters from \citep{pastorello2007b,pastorello2007a}.}
\label{fig:lcs}
\end{figure*}

\begin{figure*}
\begin{center}
\epsfig{file=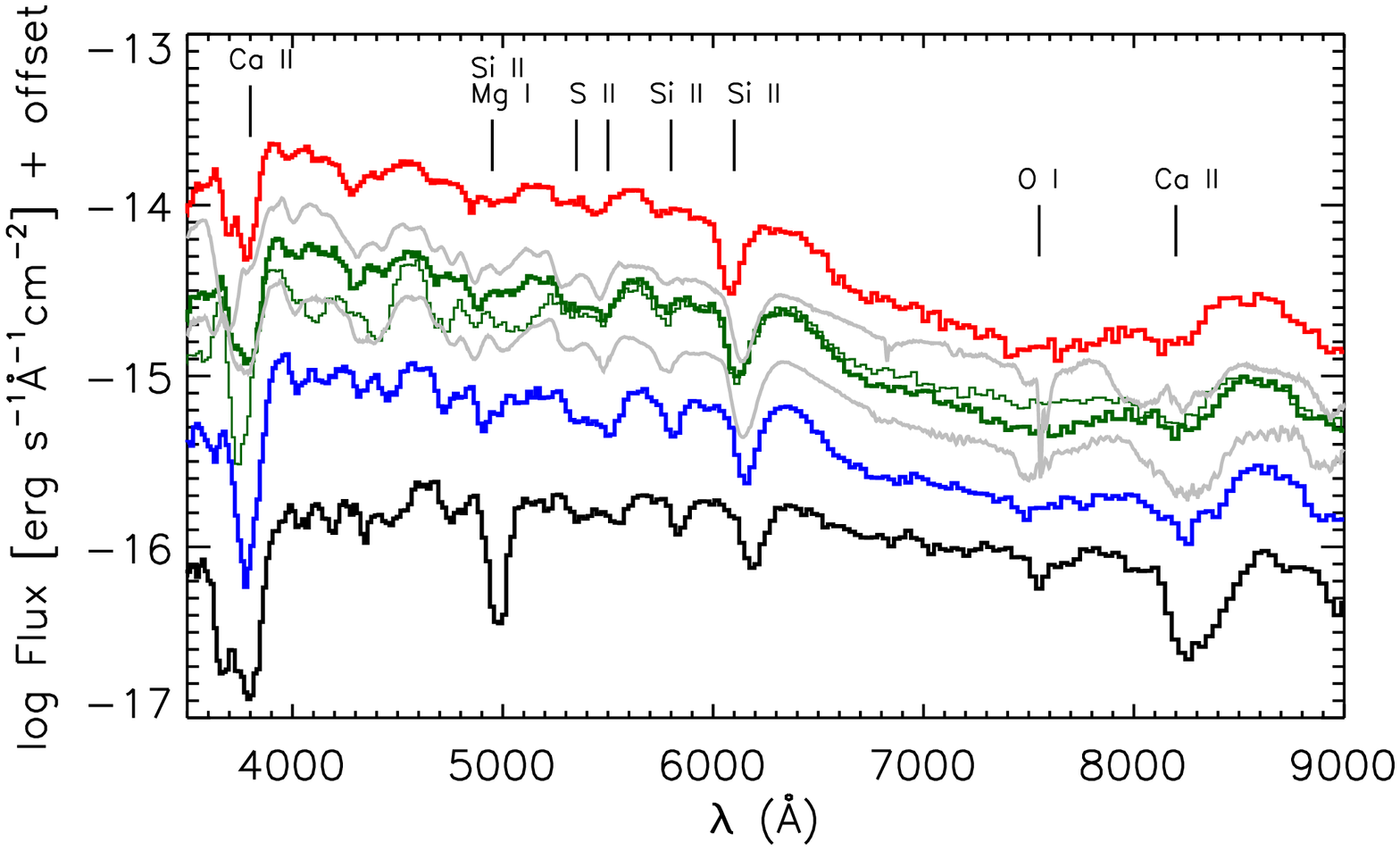,width=10cm,angle=90}
\end{center}
\caption{Computed spectra at $B$-band maximum light 
for $M_\mathrm{WD}=1.15$,
1.06, 0.97, 0.88 M$_{\odot}$ (red, green, blue and black,
respectively).
For $M_\mathrm{WD}=1.06$~M$_{\odot}$, spectra for both our {pure-C+O}
and {C+O+Ne} models are
shown (thick and thin green lines, respectively).
Observed spectra of two SNe~Ia around maximum light 
are shown for comparison: 
SN~2004eo (lower grey line; 
\citealt{pastorello2007b}) and
SN~2005cf (upper grey line; 
\citealt{garavini2007}). Arbitrary vertical offsets 
have been applied for clarity. The observed spectra are de-redshifted
and de-reddened using parameters from \cite{pastorello2007b,pastorello2007a}.}
\label{fig:spec}
\end{figure*}

In Figure~\ref{fig:lcs} we show the ultraviolet-optical-infrared
(\textit{UVOIR}) bolometric and the band-limited ($U$, $B$, $V$,
  $R$, $I$, $J$, $H$ and $K$-band) light curves from our radiative
transfer simulations. 
Maximum light spectra are shown for the same calculations in
Figure~\ref{fig:spec}.
{Observations of two SNe~Ia (SN~2005cf and 
SN~2004eo)
are shown for comparison in both figures. The $^{56}$Ni-masses
reported for these objects are significantly different but within the
range covered by our models 
($0.45{\;}\mbox{M}_{\odot}$ for SN~2004eo [\citealt{pastorello2007b}]
and $0.7{\;}\mbox{M}_{\odot}$ for SN~2005cf
[\citealt{pastorello2007a}]).}

Given the simplicity of the underlying explosion simulations, the
light curve shapes and colours are in remarkably good
agreement with observations. 
For $M_\mathrm{WD}=0.97$, 1.06 and $1.15{\;}\mbox{M}_{\odot}$
we obtain $B$-band rise times of 18--20 days, close to
observational estimates ($\sim19{\;}\mbox{days}$; \citealt{conley2006a}). The
peak colours are also close to
those observed but slightly redder in $B-V$ compared
to $M_\mathrm{Ch}$ models of similar brightness
(c.f. \citealt{hoeflich1996b,kasen2009a}).
These results differ 
from previous studies for sub-$M_\mathrm{Ch}$ models where relatively
rapid rise times (see e.g.\ 
  \citealt{hoeflich1996b}) and {\it blue} colours
  (c.f. \citealt{hoeflich1996b,nugent1997})
were found. 
Both systematic
differences arise due to  
significant amounts of $^{56}$Ni present in the outer layers of their
models (see discussion by \citealt{hoeflich1996a}). Thus, our calculations
illustrate that {\it{if}} core detonations can be triggered 
{\it{without}} producing large masses of IGEs in the outer
layers then good agreement
with observations can be obtained. A modest additional mass of IMEs or unburned
fuel ($^{12}$C, $^{16}$O or helium) in the outer ejecta due to the
triggering mechanism would have minor consequences for the spectra.

Our maximum light spectra are also in qualitatively good
agreement with observations (Figure~\ref{fig:spec}). The models all
show the characteristic Si~{\sc{ii}} $\lambda6355$ feature.
Moreover, they reproduce the sense of the observed trend whereby the strength of
the weaker Si~{\sc{ii}} feature at $\lambda5972$ relative to
$\lambda6355$
is systematically smaller in brighter events
\citep{nugent1995,bongard2006,hachinger2008a}. In agreement with
observations, the maximum light spectra show clear features
associated with other IMEs, in particular Ca and S. Our fainter models
also predict O~{\sc{i}} absorption ($\lambda7773$) as
observed in some SNe~Ia (including SN~2004eo; see
Figure~\ref{fig:spec}) but this feature
becomes weak for our brighter models, a trend also
consistent with observations \citep{nugent1995}. 
As expected from Figure~\ref{fig:abund}, there is a tendency for 
higher velocities of IME features in brighter events.
{Velocities for Si~{\sc{ii}} $\lambda6355$ measured from our spectra 
are given in
Table~\ref{tab:RT_out} and are generally compatible with those inferred from
observations (e.g. \citealt{benetti2005a}).
The Si velocity for our pure-C+O $M_\mathrm{WD}=1.06{\;}\mbox{M}_{\odot}$ model is
slightly too high for both SN~2004eo and SN~2005cf. However, this
discrepancy is small and on the scale of the differences between the
models (the observed line velocities are bracketed by those 
in our $M_\mathrm{WD}=0.97$ and $1.06{\;}\mbox{M}_{\odot}$
maximum light spectra).}

\begin{figure}
\hspace{-0.5cm}
\vspace{-0.3cm}
\epsfig{file=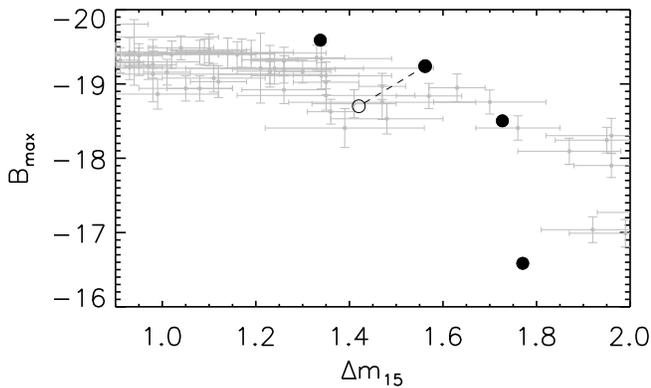,width=6cm,angle=90}
\caption{Comparison of the $B$-band 
  light curve peak magnitude ($B_\mathrm{max}$) and decline-rate parameter (${\Delta}m_{15}$)
  relation obtained from the
  models (solid circles) with observations of SNe~Ia
  (\citealt{hicken2009a}; grey crosses). SNe~Ia with distance modulus $\mu<33$ have
  been excluded. The
  open circle shows the {C+O+Ne model} 
  and is connected by a dashed line
  to the point for the {pure-C+O} model with the same $M_\mathrm{WD}$. }
\label{fig:wl}
\end{figure}

Figure~\ref{fig:wl} shows the $B$-band width-luminosity
relationship obtained from our models compared with the
properties of a sample of well-observed SNe~Ia \citep{hicken2009a}.
The models reproduce the correct systematic trend: 
brighter models have light curves which decline more slowly.
Compared to the observations the more massive models 
($M_\mathrm{WD}=1.06$ and $1.15{\;}\mbox{M}_{\odot}$) have
${\Delta}m_{15}$ larger than observed for their brightness. 
{We note, however, that ${\Delta}m_{15}$ is a particularly challenging
quantity to model precisely and systematic uncertainties in the 
radiative transfer simulations can affect this quantity
significantly. 
For example, applying different radiative
transfer codes to the 
well-known W7 model
\citep{nomoto1984a,thielemann1986a}, which is widely regarded as a good
standard for normal SNe~Ia,
yields ${\Delta}m_{15}$-values that differ by several tenths of a magnitude (see e.g. 
figure 7 of \citealt{kromer2009a}). Moreover, the values obtained are
also too large compared to those of normal SNe~Ia (e.g. {\sc{artis}}
yields ${\Delta}m_{15}\sim1.75$).
The predicted width-luminosity relation is also affected by details of
the initial WD as illustrated by our {C+O+Ne test
model}. Compared to the 
equivalent {pure-C+O model},
this model declines more slowly in B-band
and is fainter.
This effect is sufficient to move this
model to the opposite side of the observed width-luminosity relation 
(Figure~\ref{fig:wl}).}
Thus there may be potential for better agreement with observations from more detailed studies.

Finally, we note that the near-infrared light curves obtained from the
models have characteristic features which are in qualitative agreement with
observations of SNe~Ia. For the brighter models ($M_\mathrm{WD}=0.97$, 1.06,
$1.15{\;}\mbox{M}_{\odot}$), the $I$, $J$, $H$ and $K$ light curves show 
distinct secondary maxima while the faintest model $(M_\mathrm{WD}=0.88{\;}\mbox{M}_{\odot})$ has
single peaks. This is consistent with observations -- normal SNe~Ia
show double-peaked near-infrared light curves while sub-luminous
events have single maxima (e.g. \citealt{wood-vasey2008a}).
Also, the difference between the models in the $J$, $H$ and
$K$-bands at $B$-band maximum light is much smaller than in the
optical bands, consistent with the observation that SNe~Ia are better
standard candles at near-infrared wavelengths \citep{krisciunas2004a}.

\section{Discussion}

The sub-$M_\mathrm{Ch}$ model for SNe~Ia has much to
commend it. 
First, it has already been suggested by empirical modelling of
bolometric SNe~Ia light curves \citep{stritzinger2006a}
that differing ejecta
masses may be required for different SNe~Ia, a property which the
sub-$M_\mathrm{Ch}$ model may explain.
Secondly, population synthesis studies
predict large numbers
of binary systems with accreting C+O~WDs: 
\cite{ruiter2009a} estimate a Galactic rate
of $\sim10^{-3}{\;}\mbox{yr}^{-1}$ for possible explosions of
sub-$M_\mathrm{Ch}$ C+O~WDs accreting from helium-rich
companions. This is much higher than their estimate of the Galactic rate
for single-degenerate $M_\mathrm{Ch}$ explosions ($0.6-1.4\times10^{-4}{\;}\mbox{yr}^{-1}$) 
{and comparable to their estimate of the WD-WD
merger rate ($1-2\times10^{-3}{\;}\mbox{yr}^{-1}$) in systems that exceed $M_\mathrm{Ch}$.
For comparison, the observed Galactic rate of SNe~Ia is
$(4\pm2)\times10^{-3}{\;}\mbox{yr}^{-1}$ (\citealt{cappellaro99}).}

Moreover, sub-$M_\mathrm{Ch}$ models provide a simple physical
parameter which could account for the range of observed brightnesses:
the mass of the exploding C+O~WD\@. 
This parameter allows for a possible link between the typical brightness
of a SN Ia and the stellar population in which it resides.  For example,
if it can be shown that explosions in binary systems with larger $M_\mathrm{WD}$
are more often found among young stellar populations relative to their
less massive $M_\mathrm{WD}$ counterparts, the observed correlation of SN Ia
brightness with host galaxy type (e.g., \citealt{howell2001b}) might be explained.

Here we have shown that detonations of sub-$M_\mathrm{Ch}$~WDs
lead to explosions which give a reasonable match to several 
properties of SNe~Ia. Specifically, WDs with masses
between $\sim1$ and $\sim1.2{\;}\mbox{M}_{\odot}$ can reproduce a wide
range of brightness with light curves that have rise times and peak
colours in roughly the correct range. In addition, the models
reproduce the characteristic spectral features present around maximum
light and the observed trend for a higher velocity at
the inner boundary of the IME-rich layer in brighter 
SNe~Ia \citep{mazzali2007a}.
Although our {pure-C+O models} yield light curves that fade
too fast after maximum, {the models predict a width-luminosity
relation which behaves in the observed sense and we would argue that the
combination of uncertainties in radiative transfer simulations
and details of the nucleosynthesis (which is sensitive to the
progenitor composition) 
can systematically affect the decline timescale.} 
Thus there is potential for
even better agreement with improved modelling.

{There are several additional observational constraints that our
current models do not address but which should be considered in future studies.
For example, off-centre detonation might lead to 
observable effects associated with departures from spherical
symmetry (e.g. \citealt{fink2010}). Chemical inhomogeneity of the pre-explosion WD could
affect the explosive nucleosynthesis: in particular, significant gravitational
settling of $^{22}$Ne \citep{bildsten2001,garcia2008} might yield a layered ejecta structure
with a central concentration of neutron-rich isotopes as favoured by
observations (e.g. \citealt{hoeflich2004a,gerardy2007a}).}

In conclusion, detonations of naked sub-$M_\mathrm{Ch}$ C+O~WDs yield
light curves and spectra which are in qualitatively good
agreement with the observed properties of SNe~Ia. The critical
question remains whether or not realistic progenitor scenarios in
which the optical display is dominated by such an explosion can be
established: {it must involve detonation of a WD with a density
profile similar to those of our toy models without producing large
masses of high-velocity IGEs.} Any sub-$M_\mathrm{Ch}$ scenario which 
meets these criteria will likely be promising in accounting for the 
observed characteristics of SNe~Ia.

\section*{Acknowledgements}

We thank S.~Taubenberger for useful discussions and preparation of the
observational data shown in Figure~\ref{fig:wl}.
This work was partially
  supported by the Deutsche Forschungsgemeinschaft via the
  Transregional Collaborative Research Center TRR33 and the Emmy Noether Program (RO3676/1-1). Simulations were carried out on the JUGENE
  supercomputer of Forschungszentrum J\"{u}lich.

\bibliographystyle{aa}

\end{document}